\documentclass[12pt]{article}
\usepackage{amsmath,amssymb}
\usepackage{slashed}
\usepackage{graphics}
\usepackage[dvipdfm]{graphicx}
\usepackage{color}
\usepackage{url}

\makeatletter

\@addtoreset{equation}{section}
\makeatother

\begin{document}
\begin{titlepage}
\begin{flushright}
{\small IPMU13-0210}\\[-1mm]
\end{flushright}
\vspace*{1.2cm}

\begin{center}

{\Large\bf 
Higgs Pair Production at the LHC and ILC \\
from general potential
} 
\lineskip .75em
\vskip 1.5cm

\normalsize
{\large Naoyuki Haba}$^1$, 
{\large Kunio Kaneta}$^{2}$,
{\large Yukihiro Mimura}$^3$,\\
and
{\large Enkhbat Tsedenbaljir}$^{3,4}$

\vspace{1cm}

$^1${\it Graduate School of Science and Engineering,  
 Shimane University, \\ Matsue, Shimane 690-8504, Japan} \\

$^2${\it 
Kavli IPMU, University of Tokyo, Kashiwa, Chiba 277-8568, Japan
} \\

$^3${\it Department of Physics, 
 National Taiwan University,  Taipei, 10617, Taiwan (ROC)} \\

$^4${\it Institute of Physics \& Technology, Mongolian Academy of Sciences, \\
Ulaanbaatar 210651, Mongolia}

\vspace*{10mm}

{\bf Abstract}\\[5mm]
{\parbox{13cm}{\hspace{5mm}
%
Higgs cubic coupling plays a crucial role to probe
an origin of electroweak symmetry breaking.
It is expected that the cubic coupling is measured by Higgs pair production
at the LHC and ILC,
and the deviations from the standard model
can be extracted from the Higgs pair production
process, and those
can give us a hint of new physics
beyond the standard model.
We consider a general potential that achieves
the suitable electroweak symmetry breaking.
As one of the interesting models,
we suggest a non-perturbative Higgs model
in which a run-away type of potential is used.
In the model, the cross sections of pair production at the
LHC is enlarged compared to the standard model.
We also study the Higgs pair production induced by
a non-canonical kinetic term of Higgs fields
which will be important to search the pair-production at
the ILC.
}}

\end{center}

\end{titlepage}


\baselineskip=18pt

\section{Introduction}

In July 2012, the CMS/ATLAS collaborations at CERN's Large Hadron Colliders (LHC) 
reported that they had discovered a boson, which is consistent with the Higgs boson 
in the standard model (SM) \cite{Englert:1964et,Glashow:1961tr}, 
and further in 2013, they confirmed the evidence that 
it is most likely the long sought Higgs boson of the SM~\cite{Aad:2013wqa,CMS:yva,EPS-H,EPS-H-prop}. 
The Higgs boson is the last piece of the SM,
and its discovery at the LHC would complete the particle content of the theory.
All the interactions of the Higgs boson have to be investigated to see whether the 
Higgs boson has the properties expected from the SM.
While the gauge interactions among the particles have been confirmed successfully,
the other interactions in which the Higgs boson participates have not been fully explored experimentally.
This situation will change dramatically as the LHC Run II starts and even more so with the ILC experiment.

The Yukawa interactions and the Higgs self-interaction are responsible for describing
the generation of fermion masses and for inducing the electroweak symmetry breaking (EWSB), respectively.
The experimental data of the 
single Higgs production and its decay to fermions and vector bosons at the LHC
are largely consistent with the SM prediction.
The other Higgs interactions have to be probed experimentally to reveal how the EWSB occurs and whether the fermions
acquire their masses as described by the SM.

The Higgs self-coupling is one of the key parameters to investigate how the EWSB occurs~\cite{Djouadi:2005gi}.
In the SM, only the quartic Higgs coupling is allowed by the electroweak gauge symmetry 
within the renormalizable Lagrangian.
The origin of the interaction has been the subject of great debate in last few decades and an inspiration to many theories.
For most of these theories, it is expected that the quartic coupling is described by
a fundamental physics which precedes the SM at a higher energy scale.
For instance, in models with supersymmetry (SUSY),
the Higgs quartic coupling is induced by the $D$-term potential.
Therefore, the quartic coupling originates from the electroweak gauge interaction,
and it is a function of the gauge couplings,
which nicely accommodates the range of the (SM-like) light Higgs 
mass \cite{Okada:1990vk,Ellis:1990nz,Haber:1990aw}.
In addition to this questions regarding the origin of the self-coupling,
in the SM, it is not clear whether the negative sign of the Higgs squared mass parameter, which triggers the EWSB, 
has a dynamical reason and why its size remains separated from the Planck scale.

Indeed, these questions have led to the expectations and thereby many attempts to describe the EWSB as the result of radiatively or dynamically induced mechanisms. 
For instance, in SUSY extension of the SM, it is well-known that the symmetry breaking  can be induced radiatively
due to the large top quark Yukawa coupling even the high scale initial value for the Higgs mass parameter is positive.
On the other hand, in a model with dynamically induced symmetry breaking,
the Higgs self-interaction can be quite different form the SM.
In other possibilities, a new physics related to the EWSB appears at TeV scale where
the Higgs self-interaction is modified from its SM value.
In such a sense, it is important to probe the Higgs self-interaction,
which governs the essence how the Higgs boson acquires a vacuum expectation value (VEV).

After expanding the Higgs potential around the Higgs VEV, the physical Higgs particle acquires a cubic self coupling.
It is expected that the cubic coupling constant, while challenging, can be measured by Higgs pair production 
at the LHC and ILC \cite{Papaefstathiou:2012qe,Baglio:2012np,Gaemers:1984vw,Plehn:1996wb,Baur:2002qd,Fujii:2013lba}. 
The measurement of the cubic Higgs coupling provides 
important hints for the Higgs self-interaction which stabilizes 
the Higgs potential \cite{Dolan:2012ac,Contino:2010mh,LopezVal:2009qy,Djouadi:1996hp,Belyaev:1999mx,Heng:2013wia}.
The deviation of the Higgs cubic coupling from the SM can be parameterized in a model-independent way
by considering a general potential as in Refs~\cite{Boudjema:1995cb,Chivukula:1993ng,He:2002qi}.
Such a general potential may be either generated by a loop level or can be constructed
from a non-perturbative model.
At the LHC, if there is a negative contribution to the Higgs cubic coupling from these consideration,
the Higgs pair production rate always tend to be enlarged. In that case the deviations are easier to be detected.
Therefore, it is interesting to investigate a model which can induce a negative contribution
to the cubic coupling.

It is expected that more precise measurement of the cubic coupling can be done
from the processes of pair Higgs production at the ILC, compared to the LHC \cite{Fujii:2013lba}.
The process in which the cubic coupling is probed receives contributions not only from the diagrams with the cubic coupling but also from diagrams with the gauge couplings.
In order to measure the cubic coupling one has to therefore know the dependency
of the total amplitudes on the individual couplings.
Indeed, in models where the couplings differ from the SM,
one has to guarantee that the $hVV$ coupling ($V$ stands for a massive gauge boson) 
remains the same as the one in SM.
Although the experimental data of the single Higgs production 
support that the $hVV$ coupling is consistent with the SM,
$hhVV$ coupling has no such constraint at the moment and can deviate from its SM value. This happens
whenever the kinetic term of Higgs boson is given by higher dimensional effective operators.
If the $hhVV$ coupling deviates from the SM, so does the pair Higgs production cross section
even if the cubic coupling remains the  same.
This shows that it is important to investigate how the cross section depends on both the
$hhVV$ coupling and the cubic Higgs coupling.

In this paper, 
we start from a general Higgs potential,
and investigate how the cubic Higgs coupling can be modified in general.
We show that the negative contribution from the SM to the cubic coupling
enhances the cross section of the pair Higgs production via gluon fusion 
at the LHC.
From the analysis of the general Higgs potential,
we find 
a type of potential that can induce a sizable negative contribution
to the cubic coupling,
if the potential contains a piece of repulsive effect from the origin of 
Higgs configuration.
Such a type of potential (so called runaway type potential)
can be constructed in non-perturbative models.
We also investigate the correction from the general kinetic term of the Higgs boson.
We learn how the deviation from the SM couplings are parameterized,
and we investigate the parametric dependency of the
cross sections of the pair Higgs productions at the ILC and LHC.
We also construct a non-perturbative model with SUSY 
to induce the negative contribution to the cubic Higgs coupling
and enhance the pair Higgs production cross section at the ILC.
The modification of the $hhVV$ coupling in the model is also investigated.

This paper is organized as follows:
In section 2, we formulate the cubic Higgs coupling from general Higgs potential.
In section 3, we show the calculation of the cross section of the Higgs pair production at the LHC,
and the negative contribution to the cubic Higgs coupling can enlarge the cross section.
In section 4, we study the modification of the $hhVV$ coupling from the non-canonical kinetic term
of the Higgs boson, and how it affects to the pair production of the Higgs bosons at the LHC
and ILC.
In section 5, we construct a non-perturbative model by SUSY QCD,
in which a negative contribution is induced in the cubic coupling of the physical Higgs field.
Section 6 is devoted to the summary and conclusions of this paper.

\section{The cubic Higgs coupling from the general potential}

It is important to investigate the interaction of Higgs to the other particles
and to know what dynamics makes the Higgs boson have a VEV.
In the SM, the tree-level Higgs potential is given as
\begin{equation}
V = m_H^2 |H|^2 + \lambda |H|^4.
\end{equation}
It is necessary that the squared mass $m_H^2$ is negative,
and in combination with the quartic self-interaction it forces the Higgs field to acquire the VEV.
The Yukawa couplings to fermions (especially to top quarks)
and the gauge couplings are also important
for the loop corrections of the Higgs potential.

Let us describe the Higgs potential in terms of a general function:
\begin{equation}
V = V(|H|^2).
\end{equation}
The function $V(x)$ can contain any effects from loop corrections, or any non-perturbative effects. 
Surely, due to the gauge invariance, it has to be a function of $|H|^2$ (if there is only one Higgs doublet).
In unitary gauge, 
$|H|^2$ is expressed as
\begin{equation}
|H|^2 = \frac{v^2}{2} + v h + \frac{h^2}{2},
\end{equation}
where $h$ is a physical Higgs mode and $v$ denotes the Higgs VEV ($H^0 = (v+h)/\sqrt2$).
Expanding the function $V(x)$ around the VEV, we obtain
\begin{eqnarray}
V &=& V\left(\frac{v^2}{2}\right) + V^\prime \left(\frac{v^2}{2}\right) \left(vh+ \frac{h^2}{2}\right) \nonumber \\
&&+\frac12 V^{\prime\prime} \left(\frac{v^2}{2}\right) \left(vh+ \frac{h^2}{2}\right)^2
+\frac16 V^{\prime\prime\prime} \left(\frac{v^2}{2}\right) \left(vh+ \frac{h^2}{2}\right)^3 + \cdots.
\end{eqnarray}
The stationary condition (vanishing the linear term of $h$) is
$V^\prime(v^2/2) = 0$.
The mass of the physical Higgs is obtained as
\begin{equation}
m_h^2 = v^2 V^{\prime\prime} \left(\frac{v^2}{2}\right).
\end{equation}
In order to obtain the 126 GeV Higgs mass, one requires 
$V^{\prime\prime}({v^2}/{2}) = m_h^2/v^2 = 0.26$.
In the standard model, for instance, the function $V(x)$
is $V(x)= m^2 x + \lambda x^2$ and one obtains $m_h^2 = 2 \lambda v^2$.
In this expression of the Higgs mass, 
it is not necessary to solve the stationary condition $V^\prime =0$
since we use $v= 246$ GeV as an input.

The cubic interaction of the physical Higgs can be also obtained as
\begin{eqnarray}
-{\cal L}_{hhh}
=
\frac{1}{2} \left(V^{\prime\prime} + \frac13 v^2 V^{\prime\prime\prime} \right) vh^3
=
\frac{m_h^2}{2v}  \left(1 + \frac13 v^2 \frac{V^{\prime\prime\prime}}{V^{\prime\prime}} \right) h^3.
\end{eqnarray}
The tree-level Higgs potential in SM gives
$V^{\prime\prime\prime} = 0$,
and therefore,
the modification from the tree-level SM Higgs potential can be parameterized by
\begin{equation}
C_h = \frac13 v^2 \frac{V^{\prime\prime\prime}}{V^{\prime\prime}},
\end{equation}
and the ratio of the cubic coupling is expressed as
\begin{equation}
\frac{\lambda_{hhh}}{\lambda_{hhh}^{\rm SM}} = 1+ C_h.
\end{equation}

Precisely speaking, in this formulation,
$C_h$ parameterizes the deviation from the tree-level cubic Higgs coupling in SM:
$\lambda_{hhh}^{\rm SM} = 3 m_h^2/v$.
As mentioned before, the general function $V(x)$ can contain loop effects.
One can easily evaluate the contribution from the top quark 1-loop effective potential:
\begin{equation}
V(x) = m^2 x + \lambda x^2 - \frac{3}{16\pi^2} y_t^4 x^2 \left( \ln \left(\frac{y_t^2 x}{Q^2} \right) - \frac32 \right),
\end{equation}
where $y_t$ is the top quark Yukawa coupling ($m_t = y_t v/\sqrt2$) and $Q$ is the renormalization scale.
Because $V^{\prime\prime\prime}(v^2/2) = -3y_t^4/(4\pi^2 v^2)$,
we obtain 
\begin{equation}
C_h = - \frac{m_t^4}{\pi^2 v^2 m_h^2} \simeq -0.1,
\end{equation}
for the loop correction in SM.

Let us consider the following Higgs potential as a toy example:
\begin{equation}
V = m_H^2 |H|^2 + \Lambda^{4-2a} (|H|^2)^a,
\end{equation}
where $\Lambda$ is a dimensional parameter.
The minimization condition is
\begin{equation}
m_H^2 + a \Lambda^{4-2a} x^{a-1} = 0,
\end{equation}
where $x= v^2/2$.
Therefore, if $a<0$ (namely, the potential for $m_H\to0$ has a run-away kind of behavior), 
$m_H^2$ is positive.
The Higgs mass is obtained as
\begin{equation}
m_h^2 = 2a(a-1) \Lambda^{4-2a} x^{a-1} = 2(1-a) m_H^2.
\end{equation}
One can calculate
\begin{equation}
C_h = \frac23  \frac{xV^{\prime\prime\prime}}{V^{\prime\prime}}
= \frac23 (a-2),
\end{equation}
and the correction from the standard model is specified only by the exponent $a$.
We note that the correction $C_h$ is negative for the run away-type Higgs potential ($a<0$).

As one can find from the above expression, the pair Higgs production
from the general scalar potential
can be parameterized by a single parameter $C_h$.
The expansion of the scalar potential is described in unitary gauge.
Here, we comment on the case of 't Hooft-Feynman gauge:
\begin{equation}
H = \left(
\begin{array}{c}
 \chi^+ \\
 \frac{v+h+ i\chi}{\sqrt2}
\end{array}
\right).
\end{equation}
In this case, $|H|^2 = v^2/2 + vh + h^2/2 + \chi^2/2 + \chi^+ \chi^-$.
Expanding the potential $V(|H|^2)$, we obtain that
the Nambu-Goldstone (NG) bosons $\chi$ and $\chi^\pm$ are massless 
under the stationary condition $V^\prime(v^2/2) = 0$,
and they will be eaten by the gauge bosons.
The interactions between the physical Higgs $h$ and the NG bosons are
\begin{equation}
-{\cal L} = \frac{m_h^2}{v} h \left(\frac{\chi^2}{2} + \chi^+ \chi^-\right)
+ \frac{m_h^2}{2v^2} (1+3 C_h) h^2 \left(\frac{\chi^2}{2} + \chi^+ \chi^-\right).
\end{equation}
Therefore, the single Higgs production is same as the one in the SM,
but for the pair Higgs production via longitudinal vector boson fusion 
the scattering amplitude is modified by the $C_h$ parameter from the SM.
The scattering amplitude of $\chi^+ \chi^- \to hh$
is obtained as
\begin{equation}
{\cal M} (\chi^+ \chi^- \to hh)
= \frac{m_h^2}{v^2}
\left(
 1+3C_h + \frac{3(1+C_h) m_h^2}{s- m_h^2}
 + \frac{m_h^2}{t-M_W^2}  + \frac{m_h^2}{u-M_W^2}
\right).
\end{equation}
Equivalence theorem \cite{Lee:1977eg} tells us that
this scattering amplitude is same as
 the longitudinal $WW$ scattering amplitude
 up to $O(M_W^2/s)$ correction (namely neglecting gauge coupling in $M_W^2 = g^2 v^2/4$).
One can easily verify this equivalence by calculating the amplitude in unitary gauge.
However, since the 126 GeV Higgs is not very heavy compared to the gauge boson masses,
we should calculate the amplitudes in unitary gauge without neglecting the gauge couplings,
for the numerical evaluation of cross sections.



The general scalar potential can be also specified to two-Higgs doublet model (2HDM).
The scalar potential in terms of $H_1$ and $H_2$ (whose hypercharges are $-1/2$ and $+1/2$, respectively)
is a function of $|H_1|^2$, $|H_2|^2$ and $H_1 \cdot H_2 (\equiv \epsilon_{ab} H_1^a H_2^b)$ .
The cubic Higgs coupling can be written in general similarly to one-Higgs case.
We exhibit the relevant expressions in Appendix.

\section{Higgs pair production via gluon fusion at the LHC}

The Higgs cubic coupling can be probed by pair production of the Higgs boson.
At the LHC, the dominant contribution of the pair Higgs production
is the gluon fusion process.
There are two diagrams for the pair Higgs production via the gluon fusion:
(i) $gg\to h \to hh$, (ii) $gg \to hh$ via a box diagram.  
The $gg\to h$ and $gg\to hh$ couplings are generated by triangle and quadrangle
top quark loop diagrams, respectively.
The effective coupling (neglecting the top quark momentum)
can be obtained by \cite{Hagiwara:1989xx}
\begin{equation}
{\cal L}_{\rm eff} = \frac{\alpha_s}{12 \pi} (\log H) G^a_{\mu\nu} G^{a\mu\nu}
= \frac{\alpha_s}{12 \pi} 
\left(
\frac{h}{v} - \frac{h^2}{2v^2} + \cdots
\right)
G^a_{\mu\nu} G^{a\mu\nu}.
\end{equation}
Due to the opposite signs of the effective couplings (in addition to a kinematical reason),
the cross section of the pair Higgs production at the LHC
is very small at the order of $O(10^{-3})$ compared to the single Higgs production.
Inversely speaking, this fact makes the process sensitive to any additional contributions and a good probe of new physics beyond SM.

The cross section of $pp\to hh$ can be obtained by
\begin{equation}
\sigma (pp\to hh)
= \int^1_{4m_h^2/s} d\tau \frac{d{\cal L}^{gg}}{d\tau}
\hat\sigma(gg \to hh; \hat s= \tau s),
\end{equation}
and the parton-level amplitude of $gg\to hh$ (using the effective coupling) is
\begin{equation}
{\cal M}( gg\to hh) = \frac{\alpha_s}{3\pi v^2}
\left(-1 + \frac{3m_h^2(1+C_h)}{\hat s - m_h^2}\right).\label{amp}
\end{equation}
The amplitude vanishes at $\hat s = (4+3C_h) m_h^2$.
From the kinematics, we integrate the parton cross section from $\hat s = 4 m_h^2$ to $s$.
One can find that the cross section of $pp\to hh$ is enhanced for $C_h<0$ as a result.
Models which give negative $C_h$ contribution are interesting since its implication at the LHC and ILC
becomes potentially more pronounced for the Higgs pair productions and, therefore, can be scrutinized in these experiments.
We note that the run-away type potential provides an example of $C_h<0$,
as mentioned before.

\begin{figure}[tbp]
 \center
  \includegraphics[width=8cm]{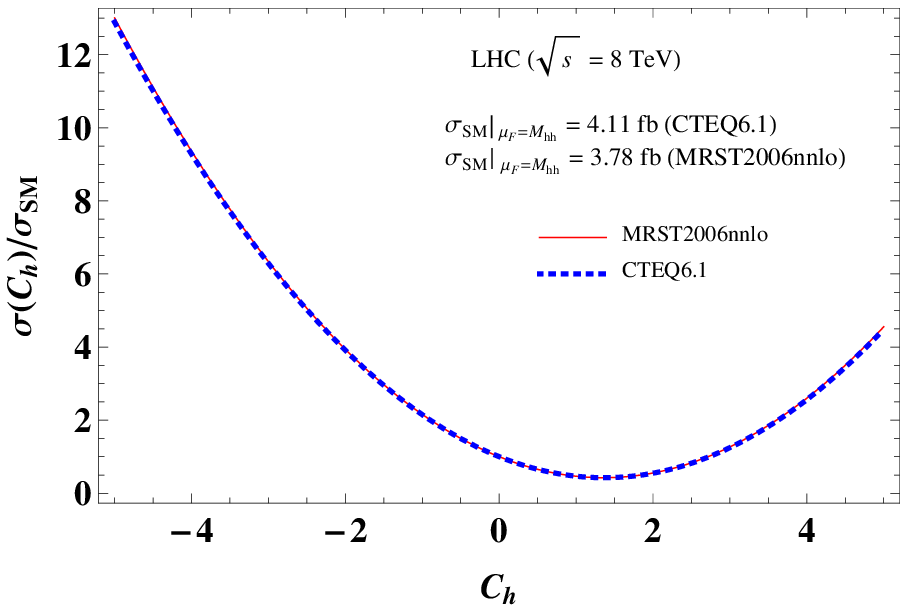}
  \includegraphics[width=8cm]{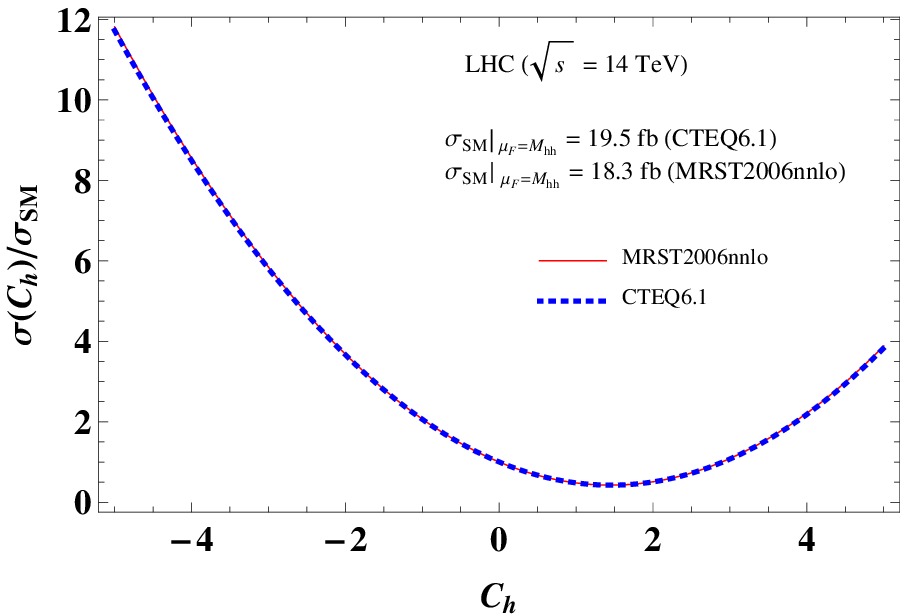}
 \caption{The ratio of cross sections $\sigma(C_h)/\sigma_{\rm SM}$
 for $gg\to hh$ at LHC, where the left and the right figures show $\sqrt{s}=8~{\rm
 TeV}$ and $\sqrt{s}=14~{\rm TeV}$ collisions, respectively. MRST2006nnlo and
 CTEQ6.1 PDF sets are used to calculate the cross section, which is
 represented by the solid (red) and the dotted (blue) lines, respectively.
}
\label{fig1}
\end{figure}

In Fig.\ref{fig1}, we show the ratio of cross sections between the
$C_h$-dependent $gg\to hh$ cross section and the SM one.
The left and right figures represent 8 TeV and 14 TeV collisions at the LHC,
respectively.
The cross sections at 8 TeV and 14 TeV at the next-to-leading order (NLO) calculation 
are 5-11 fb and 25-45 fb, respectively \cite{Papaefstathiou:2012qe,Baglio:2012np}.
The numerical numbers in the plots are given at the leading order (LO) calculation.
It is expected that the factor in the NLO/LO calculation is canceled in the $C_h$ dependence,
and thus, we show the ratio of the cross sections.
We utilized {\tt FormCalc/LoopTools}\cite{Hahn:1998yk} to
evaluate the cross sections employing MRST2006nnlo \cite{Martin:2007bv} and
CTEQ6.1 \cite{Stump:2003yu} PDF sets.
The renormalization and factorization scales are set to be equally $\mu_F$, and we take $\mu_F=M_{hh}$ where $M_{hh}$ is the invariant mass of the Higgs pair.
As a characteristic feature of the amplitude (\ref{amp}), one can find
that negative $C_h$ enhances the production cross section
compared to positive $C_h$ in the figure.\footnote{
The discovery potential for pair Higgs production at the LHC is studied in
Refs.\cite{Papaefstathiou:2012qe}. Promising channels at a large luminosity phase of the LHC
are $hh\to b\bar b W^-W^+$, $b\bar b\gamma\gamma$ and $b\bar b\tau^+\tau^-$.
}
Note that $C_h\sim 1.5$ gives the minimum value for the cross section.

\begin{figure}[tbp]
 \center
 \includegraphics[width=8cm]{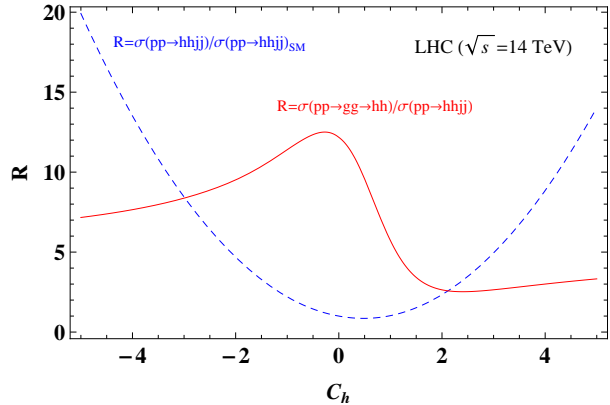}
\caption{The ratios of cross sections $R$ in which $R=\sigma(pp\to
 hhjj)/\sigma(pp\to hhjj)_{\rm SM}$ and $R=\sigma(pp\to gg\to
 hh)/\sigma(pp\to hhjj)$ are shown by dashed (blue) and solid (red)
 lines, respectively. The cross section of $pp\to hhjj$ in SM is given
 by $\sigma(pp\to hhjj)_{\rm SM}=1.6$ fb at 14 TeV LHC.}
\label{fig1-1}
\end{figure}

We comment on the Higgsstrahlung process $q\bar q \to V^* \to V hh$
and $WW$ fusion process.
At the LHC, the processes are subdominant and the cross section is an order of magnitude smaller than
the gluon fusion process in SM, where both the processes give the cross
section $\sigma(pp\to hhjj)=1.6$ fb at 14 TeV LHC.
However, if the Higgs cubic coupling is modified, these processes should
be affected.
Figure \ref{fig1-1} shows the ratios of cross sections denoted by $R$, where
$R=\sigma(pp\to hhjj)/\sigma(pp\to hhjj)_{\rm SM}$ for $pp\to hhjj$
process denoted by dashed (blue) line. In the figure, we can see that $C_h\neq 0$ can enhance the cross section.
On the other hand, when we take $R=\sigma(pp\to gg\to hh)/\sigma(pp\to
hhjj)$ for various $C_h$, which is denoted by solid (red) line in the
figure, one can find that the Higgsstrahlung and vector boson fusion
processes are subdominant compared to the gluon fusion process even if $C_h\neq 0$.

\section{Contribution from the non-canonical kinetic term of the Higgs boson}


At the ILC, the Higgsstrahlung process $e^+e^- \to Z^* \to Z hh$
and $WW$ fusion process $e^+ e^- \to WW^*\nu\bar \nu \to hh \nu\bar\nu$
 are expected to be important in probing the cubic coupling.
In particular, the Higgs cubic coupling can be measured
using the $WW$ fusion process~\cite{Djouadi:2005gi,Fujii:2013lba}.

These processes receive the contributions not just from the cubic coupling but also from the $hVV$ and $hhVV$ couplings due to the gauge interactions. 
Therefore, for more general consideration, we study cases wherein either or all of these couplings are modified from their SM values.
If the results for these processes at  the ILC  differ from the SM expectations, it is important to understand which one  of these modifications is responsible, 
since those modified Higgs-gauge boson couplings obscure the measurement of the cubic coupling.

The modification to the Higgs-gauge interactions due to the following non-canonical kinetic term has been considered\footnote{
In general, there can be a different type of operator,
\begin{equation}
(H D_\mu H^\dagger) (H^\dagger D^\mu H)
\end{equation}
 which causes the different Higgs couplings to $W$ and $Z$ bosons. 
 However, it also modifies $\rho$ parameter. Here we do not consider such a operator 
for simplicity.
 General dimension-six operators are  enumerated in Ref.\cite{Buchmuller:1985jz}. 
} in Ref.~\cite{Chivukula:1993ng}:
\begin{equation}
{\cal L}_{\rm kin} = F\left(\frac{2|H|^2}{v^2}\right) D_\mu H^\dagger D^\mu H,
\label{Lkin}
\end{equation}
where $D_\mu$ is the covariant derivative for the Higgs field.
For the convenience of the kinetic normalization, the function $F$ is defined
as $F(1) = 1$ (otherwise, the kinetic normalized field is $\sqrt{F(1)} H$).
Expanding the general kinetic function $G(x) \equiv x F(x)$, we obtain the 
$W/Z$ boson masses and coupling to the physical Higgs as
\begin{equation}
\left(M_W^2 W_\mu^+ W^{-\mu} + \frac{M_Z^2}{2} Z_\mu Z^\mu\right)
\left(1+ G^\prime(1) \frac{2h}{v} + (G^\prime (1) + 2 G^{\prime\prime}(1)) \frac{h^2}{v^2} \right).
\end{equation}
In SM, $F(x)=1$, and obviously, $G^\prime = 1$ and $G^{\prime\prime}=0$.
We denote these shifts for the couplings $hVV$ and $hhVV$ from the SM by the parameters $C_1$ and $C_2$ respectively:
\begin{equation}
1+C_1 = G^\prime(1), \qquad
1+C_2 = G^\prime(1) + 2G^{\prime\prime}(1).
\label{c1c2}
\end{equation}
%

Both $hVV$ and $hhVV$ couplings are important for the 
Higgsstrahlung and vector boson fusion processes.
As for the $hVV$ coupling, it is expected to be measured accurately
by means of the parameters relating to the single Higgs production and its decay \cite{Cheung:2013kla}
before the pair Higgs production can be observed.
In addition to this chronological reason, the $hVV$ coupling is restricted 
by oblique corrections for the precise electroweak measurements \cite{He:2002qi}, 
while $hhVV$ coupling is not.
We, therefore, fix the $hVV$ coupling in our analysis 
to its SM value $C_1 = 0$.
%
We comment that even if the single Higgs production is fully consistent with SM prediction,
it is possible that $G^\prime(1) = 1+C_1 = -1$.
In that case, however, one can redefine $h \to -h$, and the cubic $h$ coupling changes its signature,
which affects the Higgs pair production.

We note that perturbative partial-wave unitarity of $WW \to hh$
scattering \cite{Lee:1977eg} is violated unless
$(1+C_1)^2 = 1+C_2$ is satisfied.
Since we choose $C_1 = 0$, the perturbative unitarity is violated for $C_2$.
In fact, the model is described as an effective theory, 
and we expect that new particles appear at around TeV scale.

The corrections $C_1$ and $C_2$ can be generated by the
non-canonical kinetic term in Eq.(\ref{Lkin}), 
as given in Eq.(\ref{c1c2})
As a simple perturbative toy example one can consider
\begin{equation}
F(x) = 1 + a \ln x,
\end{equation}
where the coefficient $a$ contains the appropriate loop factor in the model. In this case one obtains $C_1 = a$ and $C_2 = 3a$.
On the other hand, the non-canonical kinetic function $F(x)$ may have a power like behavior if it is generated by a strong dynamics. We will give later an explicit model. Let us consider the following power function:
\begin{equation}
F(x) = x^n.
\end{equation}
In this case, one can obtain $C_1 = n$ and $C_2 = n(2n+3)$.
If $n=-2$, we have $1+C_1 = -1$. 
It is obvious that the single Higgs production is consistent with SM if $1+C_1 = -1$.
It can be understood by the (unphysical) redefinition $h\rightarrow-h$.
However, upon this change, the cubic $h$ coupling flips its sign and $C_2 =2$,
 and therefore, the cross section of pair Higgs production is modified.
This toy example can be obtained
if K\"ahler potential of the Higgs fields is given as
\begin{equation}
K = (H_1^\dagger H_1)^3 + (H_2^\dagger H_2)^3,
\end{equation}
in a SUSY model.

\subsection{LHC}

As mentioned before, the pair Higgs production via vector boson fusions and 
Higgsstrahlung are sub-dominant compared to the gluon fusion process
at the LHC.
This situation may change if the $hhVV$ couplings are modified ($C_2 \neq 0$),
so that these sub-dominant processes are enhanced.
The vector boson fusion processes can be calculated by so-called effective vector boson 
approximation \cite{Dawson:1984gx}, which can be obtained by using the amplitude 
of the longitudinal vector boson scattering to pair-Higgs bosons, as we mentioned in the previous section.
While the approximation is illustrative and easier to derive than the exact treatment, it is not particularly good
due to the fact that the self-coupling of 126 GeV Higgs boson is not so strong and gauge couplings cannot be
neglected. Therefore we use MadGraph 5 \cite{Maltoni:2002qb,Alwall:2011uj}
 for our numerical calculation which is essentially equivalent to the exact treatment. The disagreements we have obtained agrees well with the comparative study reported in Ref.~\cite{Boudjema:1995cb}.

\begin{figure}[tbp]
 \center
  \includegraphics[width=8cm]{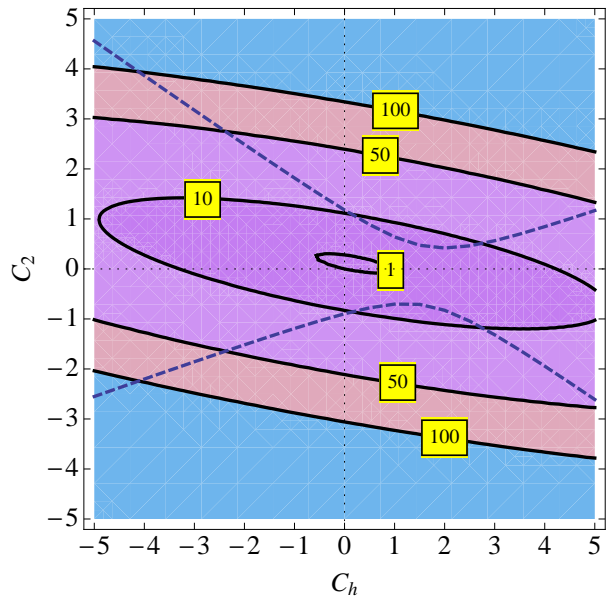}
 \caption{The Higgs pair production $pp\rightarrow hh jj$ enhancements are shown as a contour plot in the $C_h$--$C_2$ plane for the 14~TeV run at the LHC. The enhancements factors are shown as numerical labels. The dashed line shows when the process becomes equal to the leading Higgs pair production via the gluon fusion.
}
\label{fig2}
\end{figure}

We have scanned the cross--section for the process $p p\rightarrow hhjj$ by the parameters $C_h$ and $C_2$. The results are shown in Fig.\ref{fig2}. As we see deviations from the SM can be quite large. In the case for the canonical kinetic term, i.e. $C_2=0$, the enhancements are appreciable only at very large deviations at $C_h=-3$ or $4$. On the other hand, the rate is more sensitive to the changes in $C_2$ as relatively smaller values for the parameter $C_2$ lead to much more enhanced deviations compared to $C_h$.

Here we briefly note on the process $q\bar q \to hh$ which is induced at loop level. As for the SM, the rate is subleading compared to the leading gluon fusion process due to the fact that it is induced by weak interactions. We expect this remain the same even the vertices $hWW$ and $hhWW$ are modified. In the SM the unitarity for the process $WW\rightarrow hh$ is granted by the cancellation among the $s$--channel diagrams where $hWW$ and $hhWW$ are related. This is lost in the presence of nonzero $C_2$ indicating that a new physics is near by in the TeV range as we have mentioned. Therefore one should treat this as an effective operator of the form $|H|^2\bar q \slashed \partial q$ which has a corresponding counter term. At large values of $C_2$ the effect may become important. In this work we do not attempt a thorough analysis for this operator and ignore its effect.


The enhancements in the Higgs pair production at the LHC due to 
the changes in $C_2$ and $C_h$ couplings may be as large as factor of $50$ 
it is very challenging to detect them as they are still more than the order of 
magnitude below the single Higgs production. Therefore these deviations still require very high luminosity.

\subsection{ILC}

There are two processes for the pair Higgs production at the ILC,
$e^+ e^- \to Z^* \to  Zhh$ (double Higgsstrahlung) and 
$e^+ e^- \to hh\nu\bar\nu$ ($WW$ fusion process) \cite{Djouadi:2005gi}.
For the 126 GeV SM Higgs boson,
the cross section of the double Higgsstrahlung is dominant for the pair Higgs production
below $\sqrt s \simeq 1$ TeV
($\sigma(e^+ e^- \to Zhh) = 0.15$ fb at $\sqrt s = 500$ GeV).
The cross section of double Higgsstrahlung is maximized at around $\sqrt s = 600$ GeV,
and it dumps for larger $\sqrt{s}$.
The $WW$ fusion process, on the other hand, grows with larger $\sqrt{s}$,
and its cross section is comparable to the Higgsstrahlung
at around $\sqrt s = 1.2$ TeV.
Primary goal at the ILC is to refine the details of the Higgs interactions and it is expected that the Higgs cubic coupling, while challenging, can be measured.
In addition to the diagram whose contribution to the amplitude is proportional to 
the cubic coupling $(e^+ e^- \to Zh \to Zhh)$,
there are diagrams which interfere with it.
Therefore, the accuracy of the measurements of the cubic coupling using the two processes
does not directly depend on the cross sections.
In fact, the measurement of the cubic Higgs coupling is obscured by the $C_2$ contribution.
Therefore, it is important to investigate the $C_2$ and $C_h$ contribution 
to the pair Higgs production at the ILC.

\begin{figure}[tbp]
 \center
  \includegraphics[width=8cm]{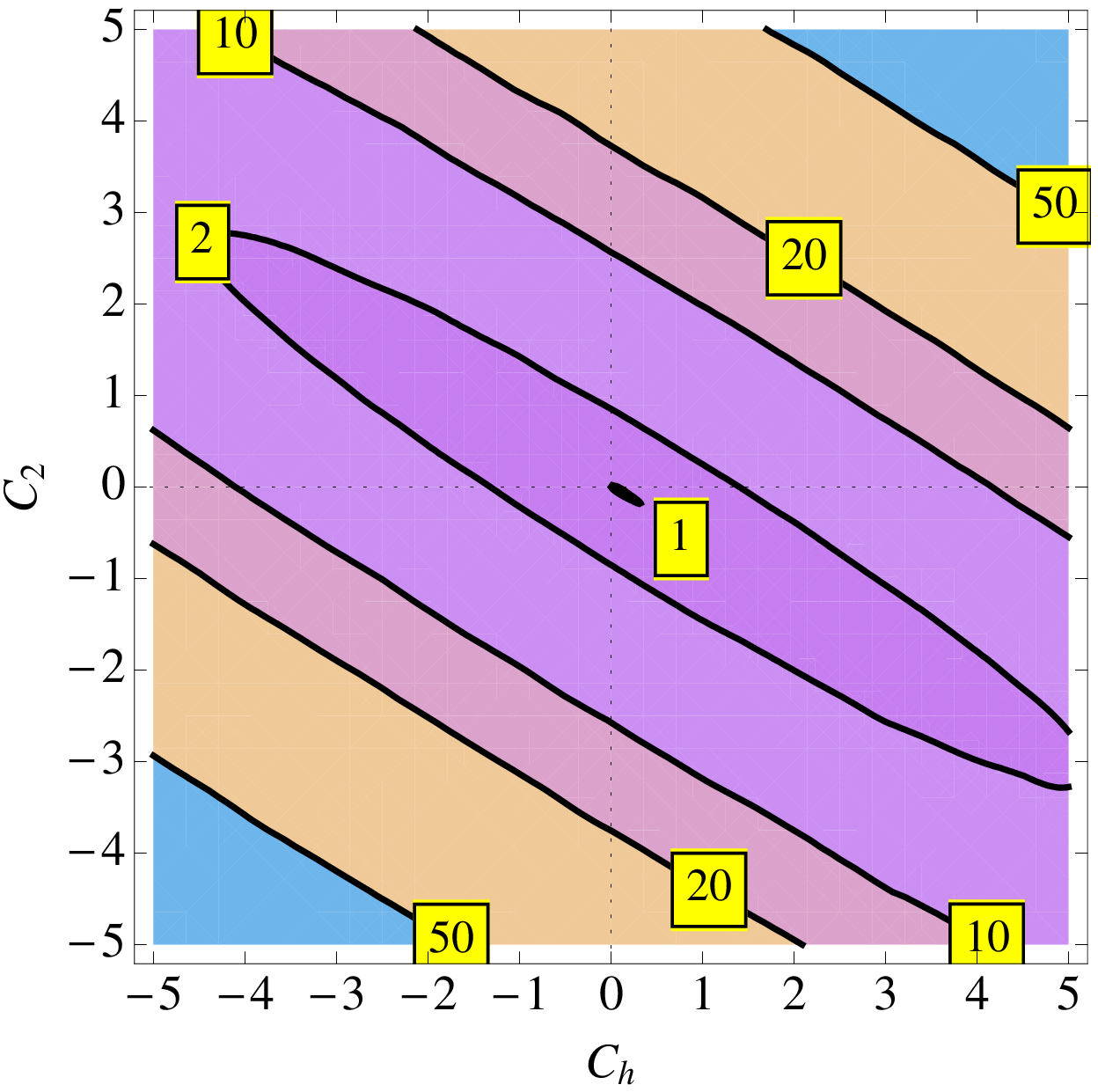}
  \includegraphics[width=8cm]{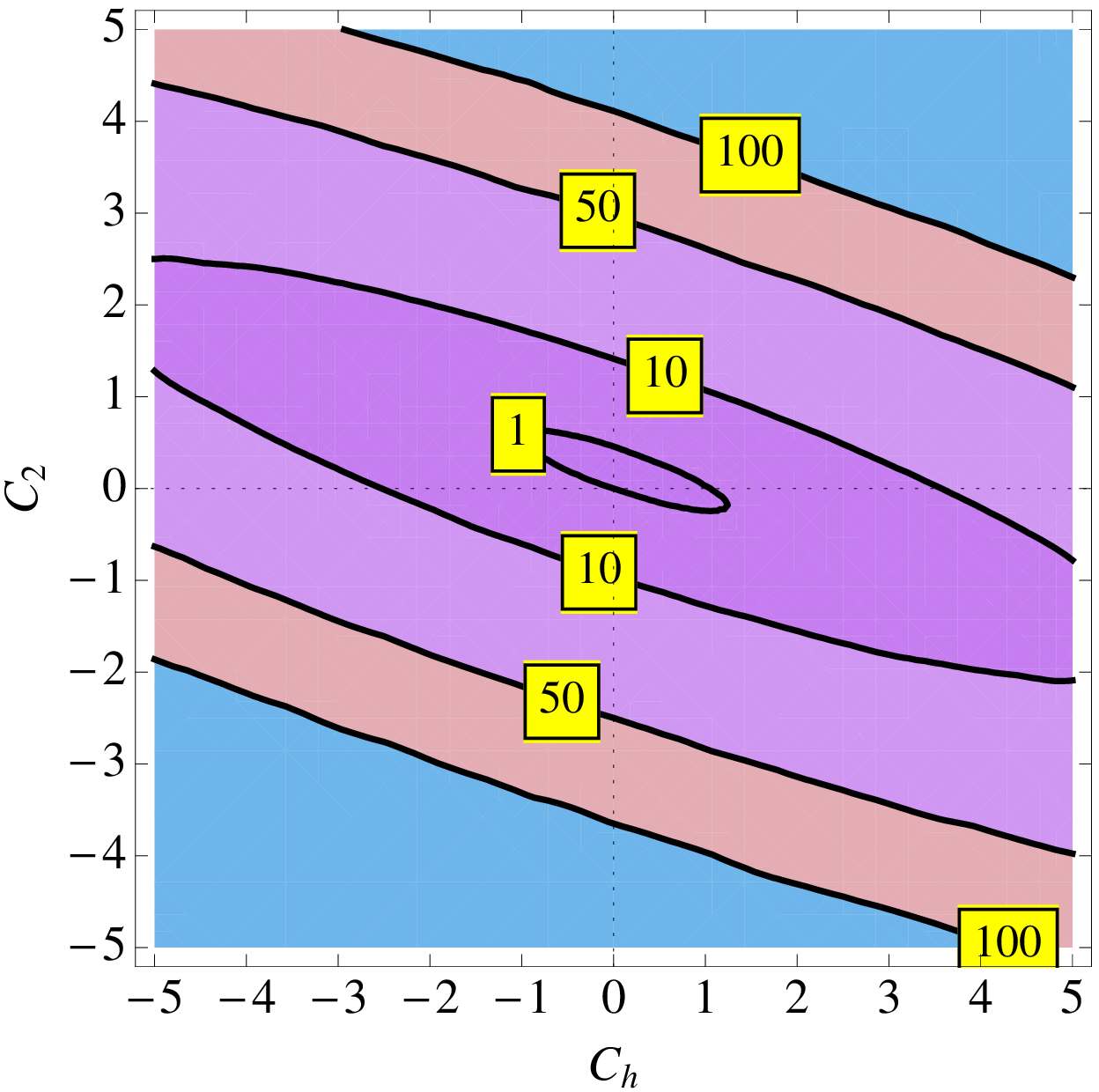}
 \caption{
The contour plots of the ratio of the cross section,
$\sigma (C_h,C_2)/\sigma(C_h=C_2=0)$
of
$e^+ e^- \to hh\nu\bar\nu$.
Left ($\sqrt s = 500$ GeV), and right ($\sqrt s = 1$ TeV).
}
\label{fig3}
\end{figure}

Similar to the LHC case, the cross section of $e^+ e^- \to hh \nu\bar\nu$
via $WW$ fusion process is calculated by using MadGraph 5
as a function of $C_2$ and $C_h$.
In Fig.\ref{fig3}, the results are shown for $\sqrt{s}=500$~GeV (right) and $1$~TeV (left) at the ILC compared to the SM expectation. 
In the first case, the enhancement is of the order of one or higher is possible in large values of $C_2$ and $C_h$ with both having the same signs. On the other hand, for the latter case the effect of $C_2$ can be dramatic with an enhancement at the level of $\sim 50$ starting from $C_2\simeq-2$ even at $C_h\simeq0$.

\begin{figure}[tbp]
 \center
  \includegraphics[width=8cm]{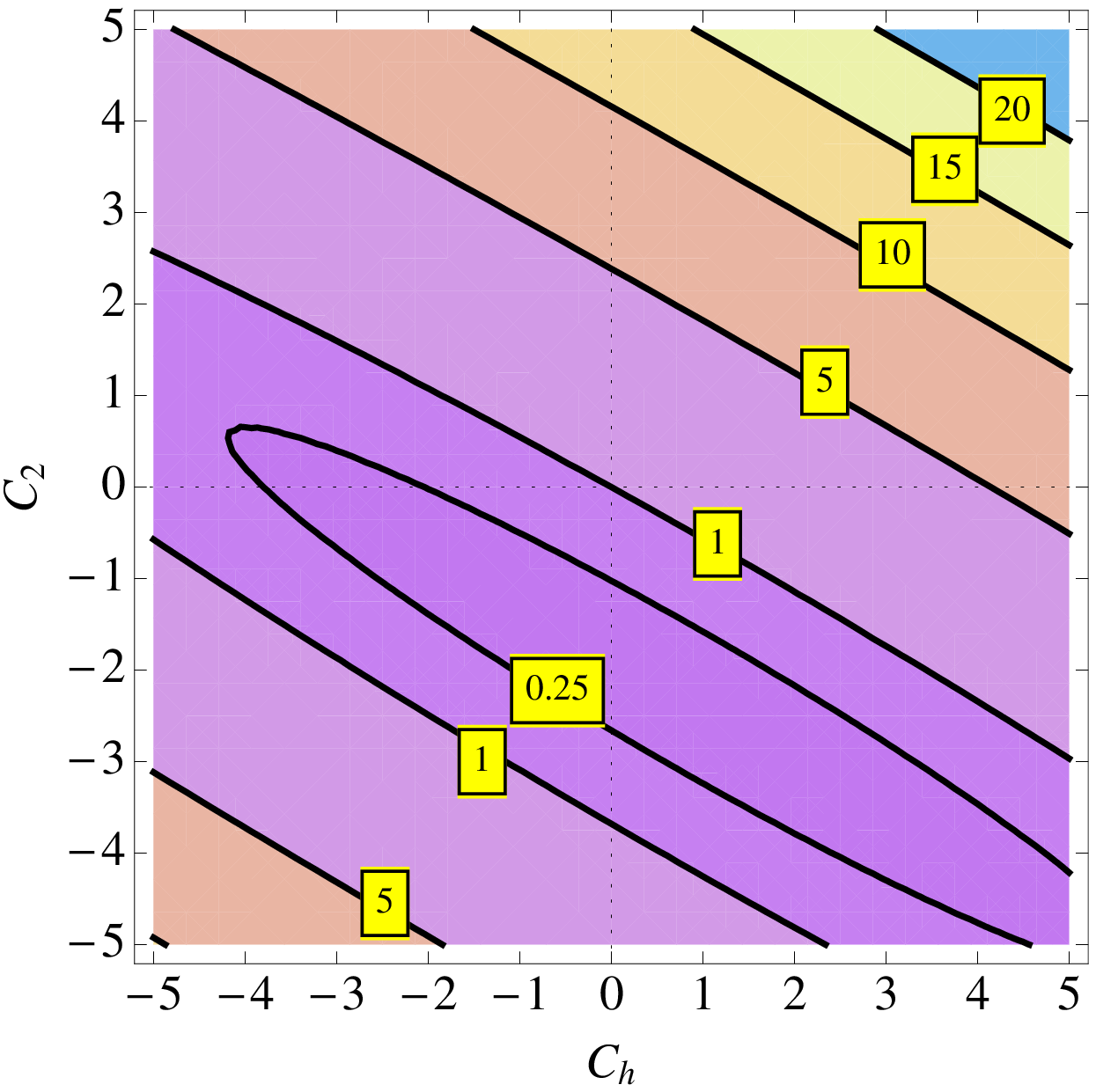}
  \includegraphics[width=8cm]{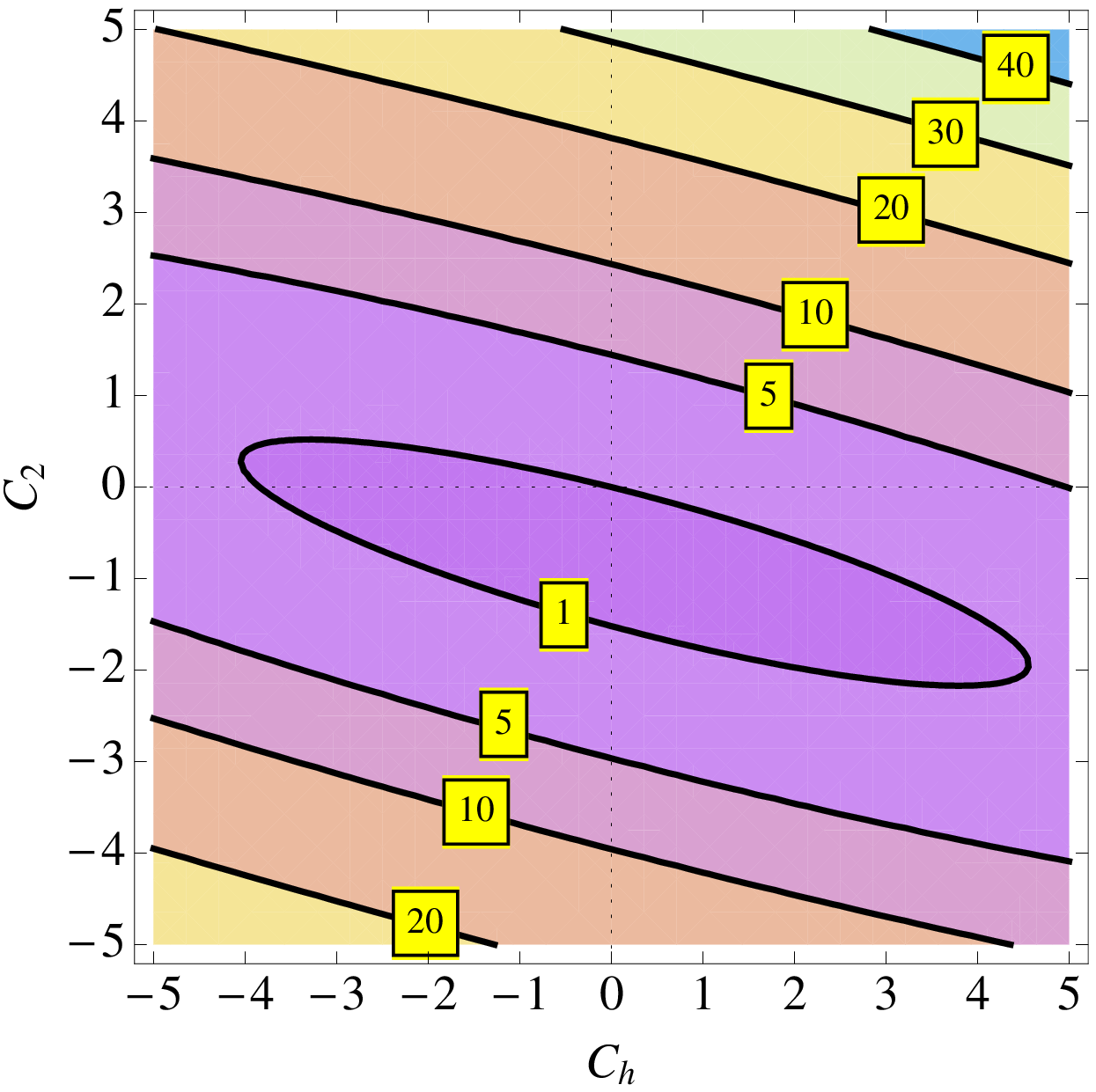}
 \caption{
The count our plots of the ratio of the cross section,
$\sigma (C_h,C_2)/\sigma(C_h=C_2=0)$
of
$e^+ e^- \to Zhh$.
Left ($\sqrt s = 500$ GeV), and right ($\sqrt s = 1$ TeV).
}
\label{fig4}
\end{figure}

In Fig.\ref{fig4}, the rates of the Higgsstrahlung process is plotted relative to the SM result. The effect is milder compared to the $WW$ fusion for both center of mass energies. This does not mean that it is more important to consider the former since their simultaneous measurement compliment each other in entangling the interference which obscures the Higgs self coupling determination.

\begin{figure}[tbp]
 \center
  \includegraphics[width=8cm]{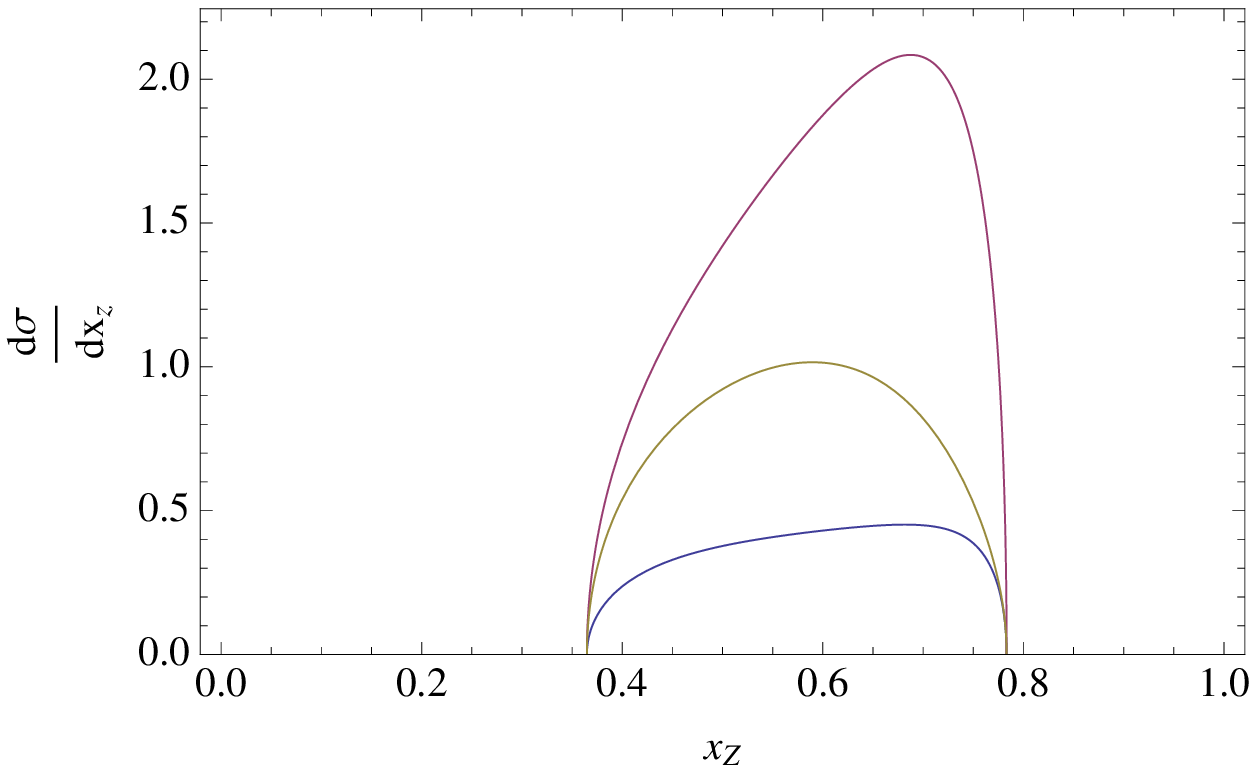}
  \includegraphics[width=8cm]{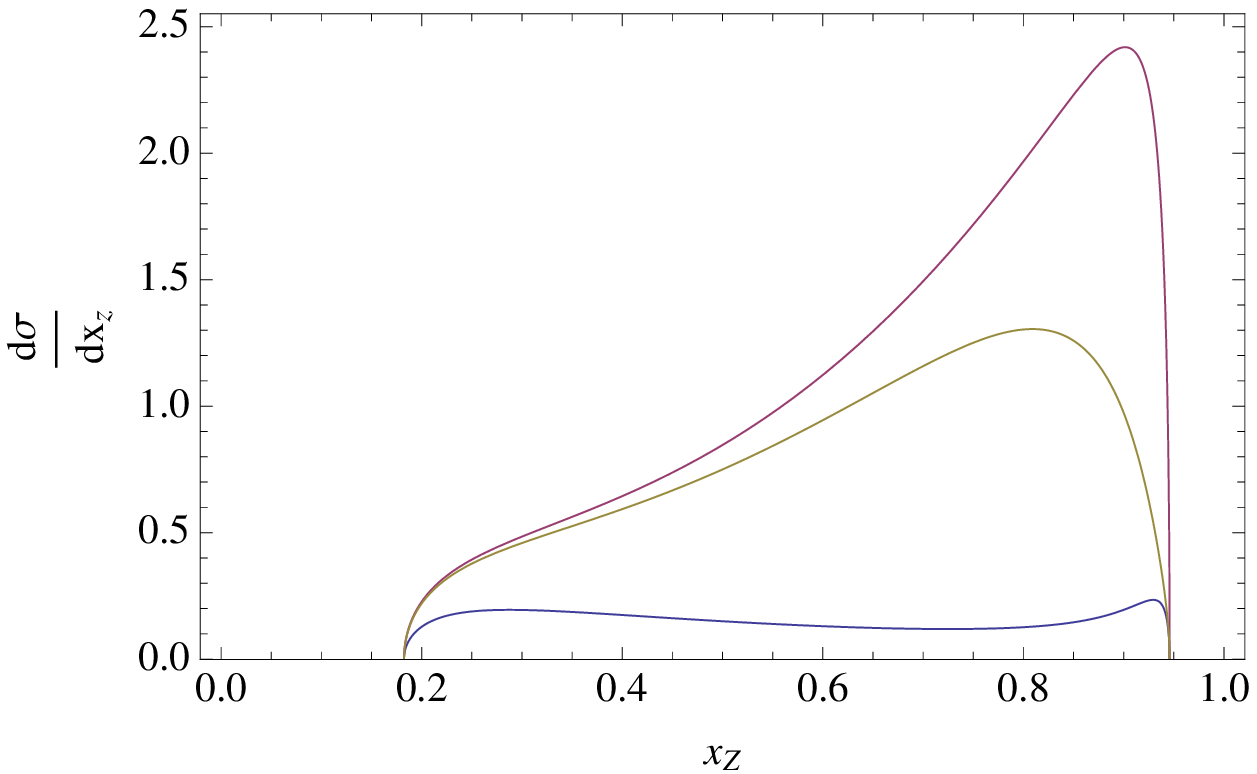}
 \caption{
  The differential cross section (in fb) of $e^+ e^- \to Zhh$.
  Left ($\sqrt s = 500$ GeV), and right ($\sqrt s = 1$ TeV).
  $x_Z$ is a scaled energy of $Z$ boson in the final state : $x_Z = 2 E_Z/\sqrt s$.
  $(C_h,C_2) = (0,0),(-2,2),(0,2)$ from below to top in each graph.
}
\label{fig5}
\end{figure}

The merit of the ILC compared to the LHC
is that the center of mass energy of $e^+ e^-$ is fixed.
The energy distribution of the final states can be used 
as a clear signal to probe the model parameters.
In fact, in addition to the cross section, 
the shape of the energy distribution of $Z$ boson is sensitive to 
the parameters $C_h$ and $C_2$.
The explicit form of 
the differential cross section of the double Higgsstrahlung is given in Ref.\cite{Djouadi:2005gi}.
In Fig.\ref{fig5},
we show the energy distribution of $Z$ boson in the $e^+e^- \to Zhh$ process
for $\sqrt s = 500$ GeV and 1 TeV.
The scaled energy of $Z$ boson $x_Z$ is defined as $x_Z \equiv 2 E_Z/\sqrt s$.
As seen from the figure,
the non-zero $C_2$ not only enhance the total cross section,
but also change the shape of the energy distribution.
We expect that $C_2$ and $C_h$ can be measured at the ILC
if there are non-SM effects in them.

At the ILC, unlike at the LHC, the environment is much cleaner which makes even mild enhancements detectable for the Higgs pair productions relatively easy. Therefore both processes are essential for determining what kind of deviations from the SM are present.

\section{A model building}

In the previous sections,
we have considered the deviation from the SM and have parameterized them as general extension.
Therefore, it can be applied to any models (perturbative, effective theories, or non-perturbative models).
As described, the pair Higgs production can be described by three parameters $C_h,C_1$, and $C_2$
(if there is only one Higgs doublets):
\begin{equation}
-{\cal L} \supset \frac{m_h^2}{2v}(1+C_h) h^3
+ \left(M_W^2 W^+ W^- + \frac{M_Z^2}2 ZZ\right)
 \left( (1+C_1) \frac{2h}{v} + (1+C_2) \frac{h^2}{v^2} \right).
\end{equation}
At the tree-level in the SM, we have $C_h = C_1 = C_2 = 0$ which
are modified by loop corrections.
If there are new particles, the quantities can become non-zero in effective theories
by integrating the heavy fields.

As we have explained in Section 3, 
the pair Higgs production at the LHC is enhanced if $C_h < 0$.
Therefore, it is interesting to build a model
in which the cubic Higgs coupling has negative contribution compared with SM.
Such a situation can be realized if the potential is run-away type non-perturbative behavior.
Indeed, the instanton effects can induce the run-away potential in SUSY $SU(N)$ QCD with $N_f$ flavor 
model for $N> N_f$ \cite{Affleck:1983rr}.
In the model, thus, 
the symmetry breaking occurs due to the non-perturbative effects of 
SUSY gauge theories \cite{D'Hoker:1996cs}.

The symmetry of the SUSY QCD is $SU(N) \times SU(N_f)_L \times SU(N_f)_R \times U(1)_B$
with chiral fields representations:
\begin{equation}
Q : ({\bf N}, {\bf N_f}, 1), \quad \bar Q : ({\bar {\bf N}},1,{\bf N_f}).
\end{equation}
The non-perturbative superpotential is generated by instanton effects \cite{Affleck:1983rr}:
\begin{equation}
W_{np} = \frac{\Lambda_0^{3+\frac{2N_f}{N-N_f}}}{({\rm det}\, \bar Q Q)^{\frac{1}{N-N_f}}},
\end{equation}
where $\Lambda_0$ is a non-perturbative scale.
We consider a case where $N_f=2$.
Suppose that $SU(N_f)_L$ is the weak gauge symmetry,
and the $U(1)$ subgroup of $SU(N_f)_R \times U(1)_B$ is the hypercharge symmetry.
(Then $N$ has to be an even number to eliminate $SU(2)_L$ anomaly.)

The composite field $\bar Q Q$, which is a moduli field
of the SUSY QCD,
 can be identified
as a Higgs bidoublet.
\begin{equation}
\bar\Lambda H_1^a = \bar Q_1 Q^a, \quad \bar\Lambda H_2^a = \bar Q_2 Q^a,
\end{equation}
where $\bar\Lambda$ is a composite scale.
Since ${\rm det}\, \bar Q Q = \bar \Lambda^2 H_1 \cdot H_2$,
the non-perturbative superpotential can be written
as \cite{Haba:2004bz,Haba:2012zt}
\begin{equation}
W_{np} = \frac{\Lambda^{3+2\alpha}}{(H_1 \cdot H_2)^\kappa},
\end{equation}
where $\kappa=1/(N-2)$, and $\Lambda^{3+2\kappa} = 
\Lambda_0^{3+2\kappa} (\Lambda_0/\bar\Lambda)^{2\kappa}$.

The K\"ahler potential in terms of the Higgs fields is obtained from the canonical form
\begin{equation}
{\cal L} = \int d^4\theta (Q^\dagger e^V Q + \bar Q e^{-V} \bar Q^\dagger),
\end{equation}
by integrating out the heavy gauge multiplet $V$
($e^V Q Q^\dagger = \bar Q^\dagger \bar Q e^{-V}$),
or using $D$-flat condition 
$D^a = Q^\dagger T^a Q - \bar Q T^a \bar Q^\dagger=0$
(Using $QQ^\dagger = \bar Q^\dagger \bar Q$ (for $N_f < N_c$),
we obtain $(Q^\dagger Q)^2 = Q^\dagger \bar Q^\dagger \bar Q Q = H^\dagger H$) \cite{Affleck:1983rr}:
\begin{equation}
K = 2 \bar \Lambda \, {\rm tr}\, \sqrt{H^\dagger H},
\end{equation}
where $H$ is a $2\times 2$ matrix ($H_i^a$), which contains two $SU(2)_L$ doublets.
Rewriting the K\"ahler potential in terms of $H_1$ and $H_2$, we obtain\footnote{
Formally, $H = H_i^a$,
and $H^\dagger H$ is a positive definite Hermite $2\times 2$ matrix.
The trace of a square root Hermite matrix $A$
is 
\begin{equation}
{\rm Tr} \sqrt{A} = \sum_i \sqrt{a_i},
\end{equation}
where $a_i$ are eigenvalues of $A$.
The eigenvalues of $A^\dagger A$ for $2\times 2$ matrix $A$
is
\begin{equation}
\frac{{\rm Tr} A^\dagger A \pm  \sqrt{({\rm Tr} A^\dagger A)^2 - 4 (\det A^\dagger A)}}{2}
=
\left(
\frac{\sqrt{{\rm Tr} A^\dagger A + 2 \sqrt{\det A^\dagger A} }  \pm \sqrt{{\rm Tr} A^\dagger A -
 2 \sqrt{\det A^\dagger A } }}{2}
\right)^2.
\end{equation}
}
\begin{equation}
K = 2\bar\Lambda \sqrt{H_1^\dagger H_1 + H_2^\dagger H_2 + 
      2 \sqrt{(H_1 \cdot H_2)^\dagger (H_1 \cdot H_2)}}.
\end{equation}
%
The K\"ahler metric from the K\"ahler potential is given in Appendix.

Using the non-perturbative potential and the K\"ahler potential,
the scalar potential can be calculated as
\begin{equation}
V_{np} = 2 \kappa^2 \frac{\Lambda^{6+4\kappa}}{\bar \Lambda}
\frac{\sqrt{|H_1|^2 + |H_2|^2 + 2 \sqrt{|H_1 \cdot H_2|^2}}}{(|H_1 \cdot H_2|^2)^{\kappa+\frac12}}.
\end{equation}
This potential is given in the case where the non-perturbative potential is exact (in SUSY limit)
and the classical K\"ahler potential is assumed.
Just for an interest, assuming that $V_{np}$ is the only piece of the
run-away potential, 
we can obtain the correction of $C_h$ for the cubic Higgs coupling:
\begin{equation}
C_h = - \frac53 - \frac43 \kappa.
\end{equation}
Because the SUSY breaking will disturb the scalar potential, we do not insist that
this potential gives the numerical quantities of $C_h$ for the cubic Higgs coupling.
However, we expect that the instanton effects induce the run-away behavior to the potential,
and it adds a negative contribution to the cubic Higgs coupling.

The kinetic term from the K\"ahler potential can be calculated as
\begin{equation}
{\cal L}_{kin}
= 
\frac{K}{2} \partial_\mu H^* \partial^\mu H 
+ 
\frac{2}{K} \left(
 (H \partial_\mu H^*)(H^* \partial^\mu H) - (H \cdot \partial_\mu H) (H^* \cdot \partial^\mu H^*)
\right),
\label{Lkin2}
\end{equation}
where the contractions of $H$ are given as
\begin{equation}
\partial H^* \partial H = \partial H^*{}_i^a \partial H_i^a,
\quad
H \cdot \partial H = \epsilon_{ij} \epsilon_{ab} H_i^a \partial H_j^b.
\end{equation}
Denoting $\langle H_1^0 \rangle = \bar v_1$
and $\langle H_2^0 \rangle = \bar v_2$ ($\bar v_1$ and $\bar v_2$ are real and positive), 
we obtain
\begin{equation}
\frac{\langle K \rangle}{2} = \bar v_1 + \bar v_2.
\end{equation}
The kinetic term of the neutral components is obtained as,
\begin{equation}
{\cal L}_{kin}^{neutral} =
2 (\bar v_1 \partial_\mu H_1^{0*}\partial^\mu H_1^0
+ \bar v_2 \partial_\mu H_2^{0*}\partial^\mu H_2^0).
\end{equation}
The kinetic normalized fields ($h$ and $H$) are defined as
\begin{eqnarray}
\sqrt{2\bar v_1} {\rm Re}\,H_1^0 &=& v_1 + \frac1{\sqrt2} (-h \sin\alpha+ H \cos\alpha), \\
\sqrt{2\bar v_2} {\rm Re}\,H_2^0 &=& v_2 + \frac1{\sqrt2} (h \cos\alpha+ H \sin\alpha).
\end{eqnarray}
From these definitions,
we obtain 
\begin{equation}
\bar v_1^3 = \frac{v_1^2}{2}, \qquad
\bar v_2^3 = \frac{v_2^2}{2}.
\end{equation}

The gauge boson mass term is obtained by replacing the derivative to covariant derivative
in Eq.(\ref{Lkin2}).
We note that the last term in Eq.(\ref{Lkin2}) does not contribute to the gauge boson mass 
due to $H\cdot \partial H = H_1 \cdot \partial H_2 - H_2 \cdot \partial H_1 =
\partial (H_1\cdot H_2)$.
In order to extract the interaction between the physical Higgs $h$ and gauge bosons,
we pick up the real part of $H^0$ which generates the gauge boson masses:
\begin{eqnarray}
{\cal L}_{V}
&=&
\frac{g^2}{2} W^+_\mu W^{-\mu} \frac{K}{2} ((H_1^0)^2 + (H_2^0)^2)\\
&+&
\frac{g^2+g^{\prime2}}{4} Z_\mu Z^\mu \left(\frac{K}2 ((H_1^0)^2 + (H_2^0)^2) + \frac{2}{K} ((H_1^0)^2 - (H_2^0)^2)^2\right).
\end{eqnarray}
If $\langle H_1 \rangle \neq \langle H_2 \rangle$,
the $\rho$ parameter $\rho = M_W^2/(M_Z^2 \cos^2\theta_W)$ shifts from 1.
Beware of the fact that VEVs of the kinetic normalized fields satisfies 
$M_Z^2 = \frac{g^2+g^{\prime2}}{2}(v_1^2 + v_2^2)$.
The $Z$ boson mass terms and the interaction to $h$ terms can be obtained as
\begin{eqnarray}
{\cal L}_Z = 
\frac{g^2+g^{\prime2}}{2} ((H_1^0)^3 + (H_2^0)^3) Z_\mu Z^\mu
=
\frac{M_Z^2}{2} Z_\mu Z^\mu
\left(1+ 3 \frac{h}{v} \sin (\beta-\alpha) h + 3 \frac{h^2}{v^2} + \cdots \right) ,
\end{eqnarray}
where $\tan\beta = v_2/v_1$\footnote{
Contrary to the case of the MSSM,
$\tan\beta =1 $ is allowed since the potential stabilization does not originate
from $D$-term potential.
}.
It is interesting to compare this result with two-Higgs doublet model:
\begin{eqnarray}
{\cal L}_Z = 
\frac{g^2+g^{\prime2}}{2} ((H_1^0)^2 + (H_2^0)^2) Z_\mu Z^\mu
=
\frac{M_Z^2}{2} Z_\mu Z^\mu
\left(1+ 2 \frac{h}{v} \sin (\beta-\alpha) h +  \frac{h^2}{v^2} \right) .
\end{eqnarray}
For $hhZZ$ coupling, thus, we obtain $C_2^Z = 2$.
For the $h$ and $W$ boson interaction terms,
the expression is complicate to show for general $\tan\beta$, and
thus we show the case $\tan\beta=1$ in which $\rho$ parameter is 1:
\begin{eqnarray}
{\cal L}_W  
=
\frac{M_W^2}{2} W^+_\mu W^{-\mu}
\left(1+ 3 \frac{h}{v} \sin \left(\frac{\pi}{4}-\alpha\right) h + (2-\sin2\alpha) \frac{h^2}{v^2} + \cdots \right) ,
\end{eqnarray}
If we choose $C_1 = 0$ (to make the single Higgs production remain unchanged), 
we obtain $C_2^W = 8/9$.

\section{Summary and Conclusions}

The discovery of the Higgs boson opens the new era of the particle physics.
The experimental data support the prediction of the  single Higgs production rate
and decays to gauge bosons by the SM.
The gluon fusion process is the dominant mechanism for the Higgs production at the LHC,
while the vector boson fusion process is subdominant and starts to be observed in the latest analysis from both experiments.
So, the gauge and Yukawa interactions for the single Higgs modes
seem to be consistent with the SM.
The decays to fermions ($b$ and $\tau$), while
 have large errors, they are consistent with the SM predictions.
It is expected that the couplings for the single Higgs production 
can be measured more accurately for the LHC run after 2015.
In addition to the single Higgs production,
it is important to observe the pair Higgs production
in order to reveal how the electroweak symmetry occurs
by Brout-Englert-Higgs mechanism.

The cross section of the Higgs pair production via gluon fusion in SM is about 25-45 fb at the LHC.
With such a low rate, it may be observed only after the measurement of couplings for single Higgs production,
decays to gauge bosons 
and fermions become more accurate.
If the pair Higgs production rate is enlarged compared to the SM,
the process can be observed earlier.
Thus, it is interesting to investigate the models,
in which the pair production rate is enlarged.
Indeed, the production rate is enlarged if there is a negative contribution 
to the cubic Higgs coupling compared to SM.
More precise measurements of various couplings are expected
at the ILC including the cubic coupling.
If any deviation is observed, it is important to know what can disturb
 the measurement of the cubic Higgs coupling.
In the case of only one Higgs doublet and $\rho$ parameter is fixed to be 1,  
all the deviations in the Higgs pair production from the SM are described by three parameters.
One of the parameters is the cubic Higgs coupling,
and other two are the $hVV$ and $hhVV$ couplings.
Even if 
the single Higgs production data turns out to be fully consistent with the SM prediction
and therefore $hVV$ coupling is fixed to comply this fact,
there are still enough room to modify the pair Higgs production rate substantially.
%
The $hhVV$ coupling can be modified if kinetic term is extended,
and it can be related to the anomalous dimension of the Higgs field.
Therefore,
the importance is the character of the Higgs boson.
For example, if the Higgs boson is a composite field,
the $hhVV$ coupling is easily modified from SM
(but $hVV$ can be also modified naively).
We have investigated the dependency of the Higgs pair productions on the two parameters which describe
deviations of the cubic Higgs coupling and $hhVV$ coupling from the SM values.
These parameters are chosen in the case where
the $hVV$ coupling is the same as the SM, keeping in mind that it will have been measured more accurately
when the pair production starts to be observed.
It is important to observe various processes
(gluon fusion, vector boson fusion, and double Higgsstrahlung)
at the LHC and ILC,
in order to determine the three couplings.
In this paper we have exhibited the parametric dependency of those processes
on the three couplings.

The pair Higgs production rate at the LHC is enlarged
if there is a negative contribution for the deviation from SM in the cubic Higgs coupling.
The negative contribution is naturally 
generated if the Higgs potential is the quadratic mass term
plus a run-away potential,
namely a repulsive effect from origin of the Higgs configuration.
It is known that such behavior can be generated by instanton effects.
Therefore, in such system, the symmetry breaking happens by the non-perturbative effects
in gauge theories.
We construct a model
in which the run-away piece exists in the Higgs potential,
and thus the pair Higgs production rate is enlarged.
We have also included the case where 
the kinetic term of the Higgs field is modified from SM. 
In this case we have shown that the pair Higgs production can 
be enlarged at the LHC and ILC compared to SM. 
This is especially important for ILC since a factor of few enhancement would be clearly measurable.
Such modifications can be tested at the LHC and ILC
by observing the various pair Higgs production processes and if observed 
may lead us to discover a mechanism behind the electroweak symmetry breaking.


\appendix

\section{Derivation of K\"{a}hler metric}

In this Appendix,
we show the calculation of scalar potential and 
kinetic term from K\"ahler potential
\begin{equation}
K= 2 \sqrt{Z + 2 \sqrt{DD^*}},
\end{equation}
where
\begin{equation}
Z = \sum_{i=1}^4 |a_i|^2,
\qquad
D = a_1 a_4 - a_2 a_3.
\end{equation}
We will identify the K\"ahler coordinates as
\begin{equation}
\left(
\begin{array}{cc}
a_1 & a_3 \\
a_2 & a_4
\end{array}
\right)
=
\left(
\begin{array}{cc}
H_1^0 & H_2^+ \\
H_1^- & H_2^0
\end{array}
\right).
\end{equation}

We obtain the K\"ahler metric as\footnote{As a common notation to describe the K\"ahler geometry,
we denote $K_i = \partial K/\partial a_i$ for example.}
\begin{eqnarray}
K_{ij^*} &=& \frac{1}{2K} (K^2)_{ij^*} - \frac{1}{4K^3} (K^2)_i (K^2)_{j^*}.
\end{eqnarray}
where
\begin{eqnarray}
(K^2)_i &=& 4 \left(Z_i + D_i \sqrt{\frac{D^*}{D}}\right), \qquad
(K^2)_{ij^*} = 4 \left(\delta_{ij^*} + 
 \frac{D_i D^*_{j^*}}{2\sqrt{DD^*}}\right). 
\end{eqnarray}

As a formulae, for a matrix,
\begin{equation}
M_{ij} = I_{ij} + X_i \bar X_j - Y_i \bar Y_j,
\end{equation}
where $I$ is an identity matrix, 
we obtain
\begin{equation}
\det M = 1 + X_i \bar X_i - Y_i \bar Y_i
- (X_i \bar X_i)(Y_i \bar Y_i) + (X_i \bar Y_i)(\bar X_i Y_i),
\end{equation}
\begin{equation}
\bar X_i M^{-1}_{ij} X_j
= \frac1{\det M} (X_i \bar X_i - (X_i \bar X_i)(Y_i \bar Y_i) + (X_i \bar Y_i)(\bar X_i Y_i)).
\end{equation}
Choosing 
\begin{equation}
X_i = \frac{D_i}{\sqrt{2D}}, \qquad
Y_i = \frac{\sqrt2}{K} \left(Z_i + D_i \frac{D^*}{D}\right),
\end{equation}
we obtain
\begin{equation}
X_i \bar X_i = \frac{Z}{2\sqrt{DD^*}},
\quad
Y_i \bar Y_i = 1,
\quad
X_i \bar Y_i = \frac{K}{4\sqrt{D^*}}.
\end{equation}
Because $Y_i \bar Y_i= 1$,
the formulae obeys
\begin{equation}
\bar X_i M_{ij}^{-1} X_j = 1.
\end{equation}
Applying the formulae to the K\"ahler metric\footnote{
$K_{ij^*} K^{jk^*}= \delta_i^{k^*}$.},
we obtain
\begin{equation}
D_i K^{ij^*} D^*_{j^*} = K \sqrt{DD^*}.
\end{equation}
When the superpotential is a function of $D$:
\begin{equation}
W = f(D),
\end{equation}
we obtain the scalar potential
as
\begin{equation}
V = W_i K^{ij^*} W^*_{j^*} =  K \sqrt{DD^*} f^\prime (D) f^\prime (D^*).
\end{equation}

The kinetic term can be obtained using the following formulae:
\begin{eqnarray}
(\det M) \bar A_i M_{ij}^{-1} A_j
&=& (\bar A A) + (\bar A A)(\bar XX)-(\bar AX)(\bar XA)-(\bar A A)(\bar Y Y)+ (\bar AY)(\bar Y A) \nonumber\\
&&-(\bar AA)(\bar XX)(\bar YY) + (\bar AA)(\bar XY)(\bar YX)\\
&&+(\bar AX)(\bar XA)(\bar YY)-(\bar AX)(\bar XY)(\bar YA) \nonumber\\
&&+(\bar AY)(\bar XX)(\bar YA)-(\bar AY)(\bar XA)(\bar YX), \nonumber
\end{eqnarray}
where 
$(\bar AA)= \bar A_i A_i$, for example.

\section{General potential for two Higgs doublets}

In this section,
we describe the Higgs self-coupling from the general scalar potential in 2HDM.
The general scalar potential is a function\footnote{
The other SU(2) invariants are a function of $|H_1|^2$, $|H_2|^2$ and $H_1\cdot H_2$.
For example,
\begin{equation}
H_1^a H_2^b (H_1^*)_b (H_2^*)_a 
= |H_1|^2 + |H_2|^2 - |H_1 \cdot H_2|^2.
\end{equation}
} of $|H_1|^2$, $|H_2|^2$ and $H_1\cdot H_2$.

In order to make the following calculation simple,
it is convenient to define linear combinations of the Higgs doublet:
\begin{equation}
\Phi_1 = H_1 \cos\beta + \hat H_2 \sin\beta ,
\quad
\Phi_2 = -H_1 \sin\beta + \hat H_2 \cos\beta,
\end{equation}
where $\hat H = i\sigma_2 H^*$,
so that the VEV of $\Phi_2^0$ is zero by definition.
We define 
\begin{equation}
x = |\Phi_1|^2, \quad
y = |\Phi_2|^2, \quad
z = \hat\Phi_2 \cdot \Phi_1   , \quad
\bar z = \hat \Phi_1 \cdot \Phi_2  ,
\end{equation}
and the general potential is a function $V(x,y,z,\bar z)$.
The stationary conditions are
$V_x = V_z = V_{\bar z} = 0$,
where $V_x$ denotes a partial derivative by $x$ for example.
We denote
\begin{equation}
\Phi_1 = 
\left(
 \begin{array}{c}
  \frac{v+\phi_1 + i\chi}{\sqrt2} \\
  \chi^-
 \end{array}
\right),
\qquad
\Phi_2 = 
\left(
 \begin{array}{c}
  \frac{\phi_2 + iA}{\sqrt2} \\
  H^-
 \end{array}
\right)
\end{equation}
The would-be-NG bosons are $\chi$ and $\chi^-$,
and $\phi_1$, $\phi_2$, $A$ and $H^-$ are physical Higgs fields.
The $\phi_1$ and $\phi_2$ fields are mixed in this basis.
Expanding the potential
around the VEV, $\langle x \rangle = v^2/2$,
we obtain the mass term of the neutral Higgs bosons:
\begin{equation}
\frac12
(\begin{array}{cc}\phi_1 & \phi_2 \end{array})
\left(
\begin{array}{cc}
v^2 V_{xx} & \frac{v^2}2 (V_{xz}+ V_{x\bar z}) \\
\frac{v^2}2 (V_{x z} + V_{x\bar z}) & V_y + \frac14 v^2 (V_{zz} + V_{\bar z\bar z} + 2 V_{z\bar z})
\end{array}
\right)
\left(\begin{array}{c}\phi_1 \\ \phi_2 \end{array}\right).
\end{equation}
The mixing angle of $H_1^0$ and $H_2^0$ is defined as $\alpha$,
and thus, 
\begin{equation}
\left(
 \begin{array}{c}
  H \\
  h
 \end{array}
\right)
= 
\left(
 \begin{array}{cc}
  \cos\alpha & \sin\alpha\\
  -\sin\alpha & \cos\alpha
 \end{array}
\right)
\left(
 \begin{array}{c}
  \sqrt2(H_1^0 - v_1)\\
  \sqrt2(H_2^0 - v_2)
 \end{array}
\right)
=
\left(
 \begin{array}{cc}
  \cos(\beta-\alpha) & - \sin(\beta-\alpha)\\
  \sin(\beta-\alpha) & \cos(\beta-\alpha)
 \end{array}
\right)
\left(
 \begin{array}{c}
  \phi_1\\
  \phi_2
 \end{array}
\right).
\end{equation}
If $V_y$ is large, $\beta-\alpha$ mixing is closed to $\pi/2$,
and $\phi_1$ is roughly the lightest Higgs boson $h$,
and 
$m_h^2 \simeq v^2 V_{xx}$.

The mass of CP odd Higgs boson $A$
and the charged Higgs mass $m_{H^+}^2$ 
can be also obtained:
\begin{eqnarray}
m_A^2 &=& V_y  + \frac14 v^2 (-V_{zz} - V_{\bar z\bar z} + 2 V_{z\bar z}), \\
m_{H^+}^2 &=& V_y.
\end{eqnarray}

The following expressions 
are useful to 
calculate the mass spectrum from the general potential.
\begin{eqnarray}
|H_1|^2 &=& x \cos^2\beta + y\sin^2\beta - \frac12 (z+\bar z) \sin2\beta, \\
|H_2|^2 &=& x \sin^2\beta + y\cos^2\beta + \frac12 (z+\bar z) \sin2\beta, \\
H_1 \cdot H_2 &=& \frac12 (x-y) \sin2\beta + z \cos^2\beta - \bar z \sin^2\beta.
\end{eqnarray}

In the two-Higgs-doublet model,
the cubic coupling can be modified
from $m_h^2/v^2$ 
if $\cos(\beta-\alpha) \neq 0$ even in the renormalizable model.
Surely, the lightest Higgs and vector bosons $hVV$ coupling is proportional to $\sin(\beta-\alpha)$
and a sizable value of $\cos(\beta-\alpha) \neq 0$
can modify $h\to WW$ and $h \to ZZ$ decays.
If we neglect the $\cos(\beta-\alpha)$ contribution,
the modification from the cubic coupling and $hhVV$ coupling
is given by $V_{xxx}$.

The physical mass parameters are related to the second derivatives of $V$ as follows:
\begin{eqnarray}
v^2 V_{xx} &=& s^2 m_h^2 + c^2 m_H^2,\\
v^2(V_{xz}+V_{x\bar z}) &=& 2sc(m_h^2-m_H^2),\\
\frac{v^2}4 (V_{zz}+V_{\bar z\bar z} + 2 V_{z\bar z}) &=& c^2 m_h^2 + s^2 m_H^2 - m_{H^+}^2.
\end{eqnarray}
where $s= \sin(\beta-\alpha)$, and $c= \cos(\beta-\alpha)$.

The cubic $hhh$ coupling is written as
\begin{eqnarray}
\lambda_{hhh} &=& \frac{s(1+c^2)}{2v} m_h^2 
-\frac{c^2s}{v} m_{H^+}^2 + \frac{v}{4} c^3 (V_{yz}+V_{y\bar z})
+ \frac{v}2 c^2 s V_{xy} \\
&+&\frac{v^3}6 s^3 V_{xxx}
+ \frac{v^3}4 cs^2 (V_{xxz}+V_{xx\bar z})
+ \frac{v^3}8 c^2 s(V_{xzz}+V_{x\bar z\bar z}+2 V_{xz\bar z}) \nonumber\\
&&+ \frac{v^3}{48} c^3 (V_{zzz}+3V_{zz\bar z}+3 V_{z\bar z\bar z}+V_{\bar z\bar z\bar z}) . \nonumber
\end{eqnarray}
The $hhH$ coupling also effects to the pair Higgs production if $\cos(\beta-\alpha)$
is not small and $H$ is not very heavy.
The $hhH$ coupling is
\begin{eqnarray}
\lambda_{hhH} &=& \frac{c^3}{v} m_h^2 - \frac{cs^2}{2v} m_H^2 
-\frac{c(c^2-2s^2)}{v} m_{H^+}^2 - \frac{3v}{4} c^2s (V_{yz}+V_{y\bar z})
+ \frac{v}2 c(c^2-2s^2) V_{xy} \nonumber \\
&+&\frac{v^3}2 cs^2 V_{xxx}
+ \frac{v^3}4 s(2c^2-s^2) (V_{xxz}+V_{xx\bar z})
+ \frac{v^3}8 c(c^2-2s^2)(V_{xzz}+V_{x\bar z\bar z}+2 V_{xz\bar z}) \nonumber\\
&&- \frac{v^3}{16} c^2s (V_{zzz}+3V_{zz\bar z}+3 V_{z\bar z\bar z}+V_{\bar z\bar z\bar z}) . 
\end{eqnarray}



\section*{Acknowledgments}

This work is partially supported by Scientific Grant by Ministry of Education and
Science, Nos. 00293803, 20244028, 21244036, 23340070, and by the
SUHARA Memorial Foundation.
 The work of K.K. is supported by Research Fellowships of 
 the Japan Society for the Promotion of Science for Young Scientists and
 also by World Premier International Research Center Initiative (WPI Initiative), MEXT, Japan.
The work of Y.M. is supported by the Excellent Research Projects of
 National Taiwan University under grant number NTU-98R0526. 
E. T. acknowledges the support from the National Science Council of 
Taiwan under Grant No. NSC 100-2119-M-002-061.
%
We thank the authors of the first paper in Ref.~\cite{Papaefstathiou:2012qe} for providing their MadGraph implementation of the Higgs pair production and explaining the details.


\end{document}